\documentclass[11pt]{article}
\usepackage{moriond,epsfig}

\def\be{\begin{equation}}
\def\ee{\end{equation}}
\def\bea{\begin{eqnarray}}
\def\eea{\end{eqnarray}}

\begin{document}
\vspace*{2 cm}
\title{ MSSM, Msugra and the LSP at LEPII}

\author{ I.Laktineh }

\address{Institut de Physique Nucl\'eaire de Lyon}

\maketitle\abstracts{ More than one year after the end of LEPII,
many analysis activities are still going on  to translate the
negative search results of the four LEP experiments into solid
limits on cross-sections and masses of  SUSY particles. Many
 analyses based on the MSSM and Msugra  models are presented in
this paper. Preliminary results including  the limit on the mass
of the lightest supersymmetric particle (LSP) within the  RP
conservation hypothesis are also given.}

\section{INTRODUCTION}
SUSY \cite{susy}  is an appealing theory  since it solves  many  of the
Standard Model (SM)  problems while conserving  its successful
features. In SUSY each boson (fermion) of  the SM, has a new
fermionic (bosonic) partner  with the same mass. Absence of
experimental observation of partners  having the same mass led to
the conclusion that SUSY is broken.  There is no clear indication
how this is done. However, few scenarios were proposed to explain
the way  this breaking is propagated to  the electroweak scale.
In one of these scenarios, called the supergravity scenario, the breaking is mediated by gravitation interaction. The
minimal supersymmetric extension of the Standard Model inspired by
the previous scenario is called the MSSM. Although this model
contains many parameters, only few of them are relevant for SUSY
particle  searches and can be summarized by the gaugino mass
terms $ M_{i}\,\, (i=1,3)$, the scalar fermion masses $m_i$, the
trilinear coupling constants $A_i$, the ratio of the V.E.V of the
two Higgs doublets   $tan \beta$ and  the mixing Higgs parameter
$\mu$.
 The number of these parameters is reduced when the unification  relations are assumed.
  In this case, the gaugino mass terms are all identical $(m_{1/2})$  at the GUT scale as
   well as the sfermion masses $ (m_{0})$ and the trilinear coupling constants $(A)$.
 Using the Renormalization Group Equations\cite{RGE} (RGE) , the gaugino mass terms are related to each other at low energy scale by the relations $ M_1:M_2:M_3=1:1.95:6.64$ \footnote{In this case the MSSM is commonly called the constrained MSSM.}.

The Msugra  is even a more restricted model with only four
parameters:$\, m_{0}, m_{1/2} , A, tan \beta$ and the sign of $\mu$.
In this  model the electroweak breaking is induced by the SUSY
breaking  which explains the absence of  $\mu $  as a
parameter.

In the two previous SUSY models (MSSM, Msugra), the lightest
neutralino\footnote{ Neutralinos are linear combinations of supersymmetric
partners of $\gamma$, $Z$ and Higgs field neutral components.} is the lightest
supersymmetric particle {\bf (LSP)} for most of SUSY parameters. The
lightest neutralino  is therefore considered as the LSP for all
the SUSY searches within the MSSM and Msugra models studied by the
LEP experiments.

Each of the four LEP experiments (ALEPH, DELPHI, L3, OPAL)  has
accumulated  an average  luminosity  of  about $660 \,\, pb^{-1}$ at
center of mass energies going from  189 to 208 GeV. This important
luminosity  has allowed to test many phenomenological aspects of
the mentioned SUSY models. The absence of a significant deviation
with respect to the SM prediction  has then led to set limits on  the SUSY
parameters as  well  as SUSY particle masses.
\\

This paper is organized as follows. In the first section we give a
short description of the different SM backgrounds associated  to
SUSY experimental searches. In the second section   the scalar
fermion searches and their preliminary results are shown.
Chargino and neutralino searches are quoted  in the third
section. In the fourth section details concerning the LSP
mass lower  limit are given. All the analyses presented in this paper are done within the R-Parity\footnote{This parity leads to a stable LSP.} conservation(RPC) hypothesis.

\section{SM CONTRIBUTION}
\begin{figure}
\begin{center}
{\psfig{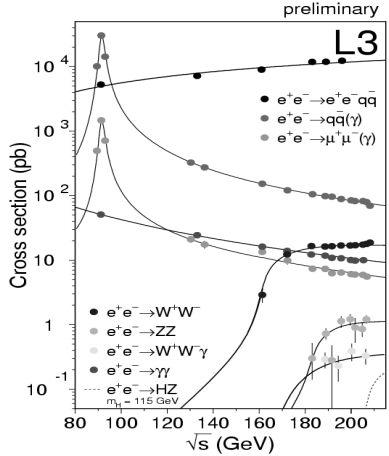}}
\end{center}
\caption{Comparison between data and SM prediction for different
physics processes.
 \label{fig:comp}}
\end{figure}

Searching for SUSY particles at LEP within RPC framework is characterized  by looking for events with missing energy. This is due to the production in the RPC
hypothesis, in the final state, of two stable LSP particles escaping
the detector.  This makes  the difference between the produced
SUSY particle mass and the LSP one $(\Delta m = m(SUSY)-m(LSP))$ an
important parameter for the experimental searches  since  the nature of
the SM background depends strongly on it. At low  $\Delta
m$, the main contribution to the background of RPC-MSSM  events
is  the two-photon physics in which the two electrons  exchange
two photons which collide giving birth to low energy particles
whereas the two electrons go undetected in the vacuum tube. At
intermediate and high values of $\Delta m$, physics processes like
2-fermion and 4-fermion final states  such as  $\,W W, \,Z Z,\, W e \nu
$ and $f \bar f$, are the dominant ones. Since the new physics
events should manifest themselves  as an excess with respect to
the SM  physics  events, the prediction of these SM events must
be under control. This is indeed the case as can be shown in figure~\ref{fig:comp} where the cross section of the different SM processes
are measured and compared to the predicted ones.

\section{SFERMION SEARCHES}
 SUSY partners of the SM fermions are scalars. There are two SUSY partners,  named left
  and right, for each SM fermion.
 Left and right  sfermions of each family have the same quantum numbers and can thus mix with each other giving  rise to new   mass eigenstates  through the following matrix:
$\left(
\begin{array}{cc}
  m_{f_{\tilde L}}& m_f (A-\mu \,f(\beta))  \\
  m_f (A-\mu \,f(\beta)  & m_{f_{\tilde R}}
\end{array}
\right)$  where $m_{f_{\tilde L}} (m_{f_{\tilde R}})$ is the mass of the
left (right) sfermion and $m_f$ is its fermion partner mass and $f= \tan(\beta)(\cot(\beta))$ for down(up)-like fermion respectively. The
new mass eigenstates can be written as: $ \tilde f_{1(2)} =
\tilde f_{L} \cos{\theta}+(-)  \tilde f_{R} \sin{\theta}\,
$ with  $\theta$ being the mixing angle. Since the mixing is
proportional to the partner fermion  mass, only mixing in the
third generation $(\tilde \tau, \tilde b,    \tilde t)$ is of
interest.
 Both sleptons and squarks, SUSY partners of leptons  and quarks respectively,
  have been searched by the  LEP experiments.
\subsection{SLEPTONS}
Assuming the same scalar mass  $m_0$ at GUT scale, masses of the
different scalar leptons at the electroweak scale can be predicted
through the RGE equations. The masses of charged left
and right  sleptons are given by:
 $$m(\tilde l_L) = m_0^2 + 0.77 M_2^2-0.27 m_Z^2 cos(2 \beta)\\ \, \, ,  \, \, m(\tilde l_R) =  m_0^2 + 0.22 M_2^2-0.23 m_Z^2 cos(2 \beta)\\$$
The scalar sneutrino mass is given by the formula: $ m(\tilde\nu) = m_0^2 + 0.77 M_2^2 +0.5 m_Z^2 cos(2 \beta).\\$
 For values of $tan \beta > 1$ \footnote{ which is the case if
Higgs negative searches are included.},  right slepton is lighter
than the left one. This determines the search
strategy of scalar leptons like  selectrons and smuons
by looking first for the right sfermions, the  more probably
accessible at LEP.
\\
\vglue .3 cm
\noindent
{\bf SMUON}:  In addition to the missing energy, the signature of scalar muon
 $\tilde \mu$ pair production is characterized by the presence in the detector of two acoplanar muons. These muons result from the decay of the
scalar muons: $\tilde \mu \to \mu +\chi^0_1$. The comparison
between data and SM prediction at the different center of mass
energies shows good agreement. LEP SUSY working group \cite{suwogr} has combined
the four experiments results to set a lower  limit on the right
smuon pair production cross-section in the plane $(m(\tilde
\mu_R),m(\tilde \chi^0_1))$. The smuon pair
production which takes place through the s-channel does not depend
on SUSY parameters directly. It depends only on smuon mass. This
allows to determine an exclusion area in the previous plane. A preliminary result shows that for
$m(\tilde \chi^0_1)$ less  than $40\,  GeV$, LEP experiments exclude smuon mass up to
 $96.4 \,  GeV$ at $ 95$ \% CL.

\vglue .3 cm
\noindent
{\bf STAU:} Scalar tau $\tilde \tau$ pair production at LEP is searched by looking for two
 low multiplicity jets corresponding to the two produced taus  from
 the stau decay:$\tilde \tau \to \tau +\chi^0_1$.
 The notion of jet is extended here to take into consideration the
  leptonic decay of the tau. The difference with respect to the scalar
  muon case comes from the possibility  of important mixing between left
  and right staus which may result into light stau ($\tilde \tau_1$).
  The mixing effect can be very important with respect to the pair
  production cross-section. Indeed for a mixing angle of $\theta_\tau= 46^\circ$,
  $\tilde \tau_1$ decouples from $Z$  leading to  a minimal scalar tau pair
  production. As in the scalar muon case, the absence of significant deviation  with
  respect to  SM prediction has been  translated by the LEP SUSY working group into a lower
  limit on the $\tilde \tau_1$ mass in the  decoupling mixing scenario.
  For $m(\chi^0_1) < 40\,  GeV$
    LEP excludes scalar tau mass up to $  87.1 \, GeV$ at $ 95 \% $CL.
\vglue .3 cm
\noindent
{\bf  SELECTRON, SNEUTRINO:} The scalar electron $\tilde e$  pair production  proceeds not only through the
s-channel  as for  the previous sleptons but also through the
t-channel. In this  case the dependence on SUSY parameters is
direct through the $e \tilde e \tilde \chi^0_1 $ coupling. The
lowest pair production cross-sections are obtained  for $tan \beta
\approx \sqrt{2}$ and negative values of $\mu$(-50 to -200).  The
absence of data excess with respect to  SM prediction in the 
$\tilde
 e_R$ pair production, characterized by two acoplanar electrons can
be used, as before, to set a lower limit on the selectron mass.
As for the other sleptons, when the right selectron is degenerate
in mass with the lightest neutralino, the detection efficiency is
very low  and $\tilde e$ pair production may not
be experimentally accessible through the two acoplanar electrons
search. However, in contrast with the other scalar leptons, selectrons
production is not restricted to $ \tilde e_R \tilde e_R$ or  $
\tilde e_L \tilde e_L$ but also includes, through the t-channel,
the $ \tilde e_R \tilde e_L$  production. This additional
contribution can be very helpful in the degenerate scenario as
long as  the  $ \tilde e_R \tilde e_L$ is kinematically
accessible. In this case the two acoplanar electrons search can be
replaced by  looking for a single electron  coming from the $\tilde e_L$ decay and  possibly accompanied
  by another soft electron resulting from the $\tilde e_R$ decay.

\begin{figure}

\begin{center}
{\psfig{figure=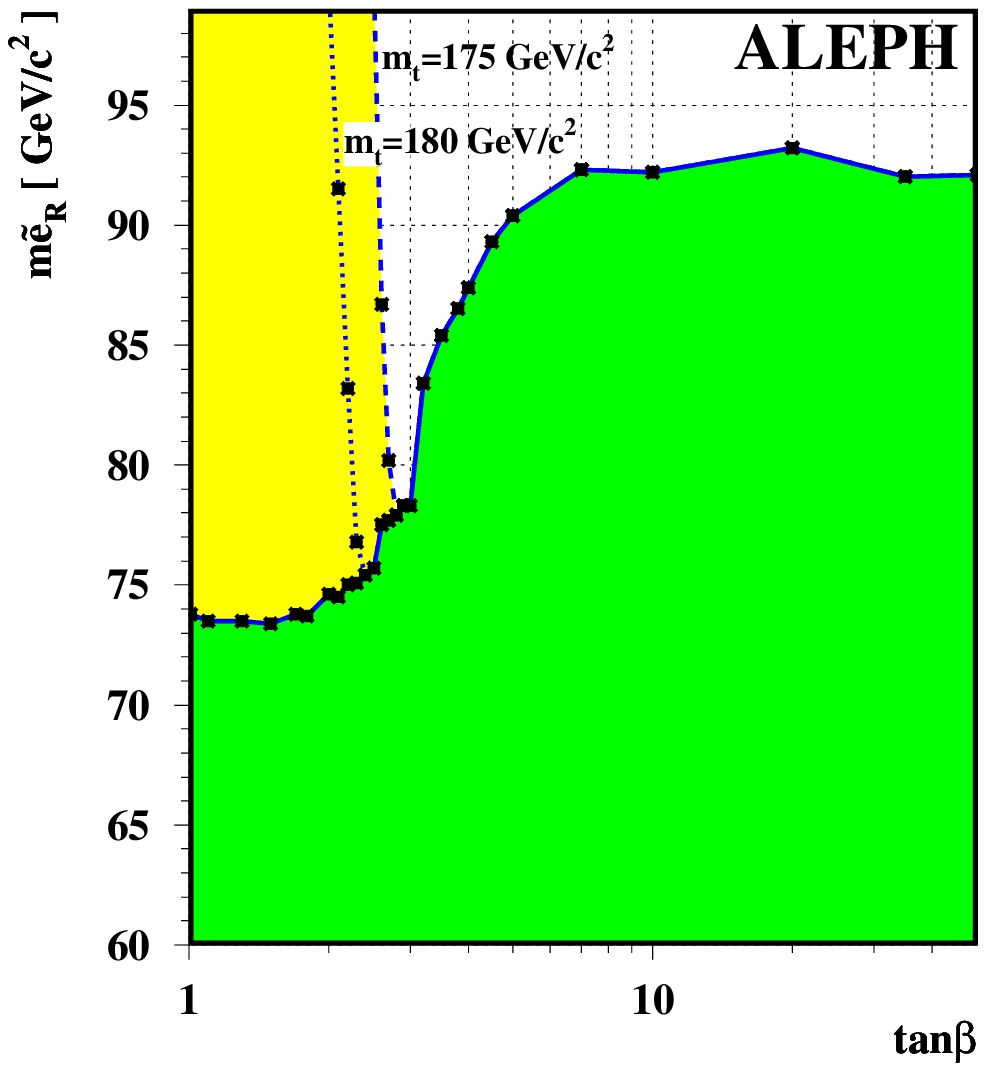,width=5cm,height=5cm}}
{\psfig{figure=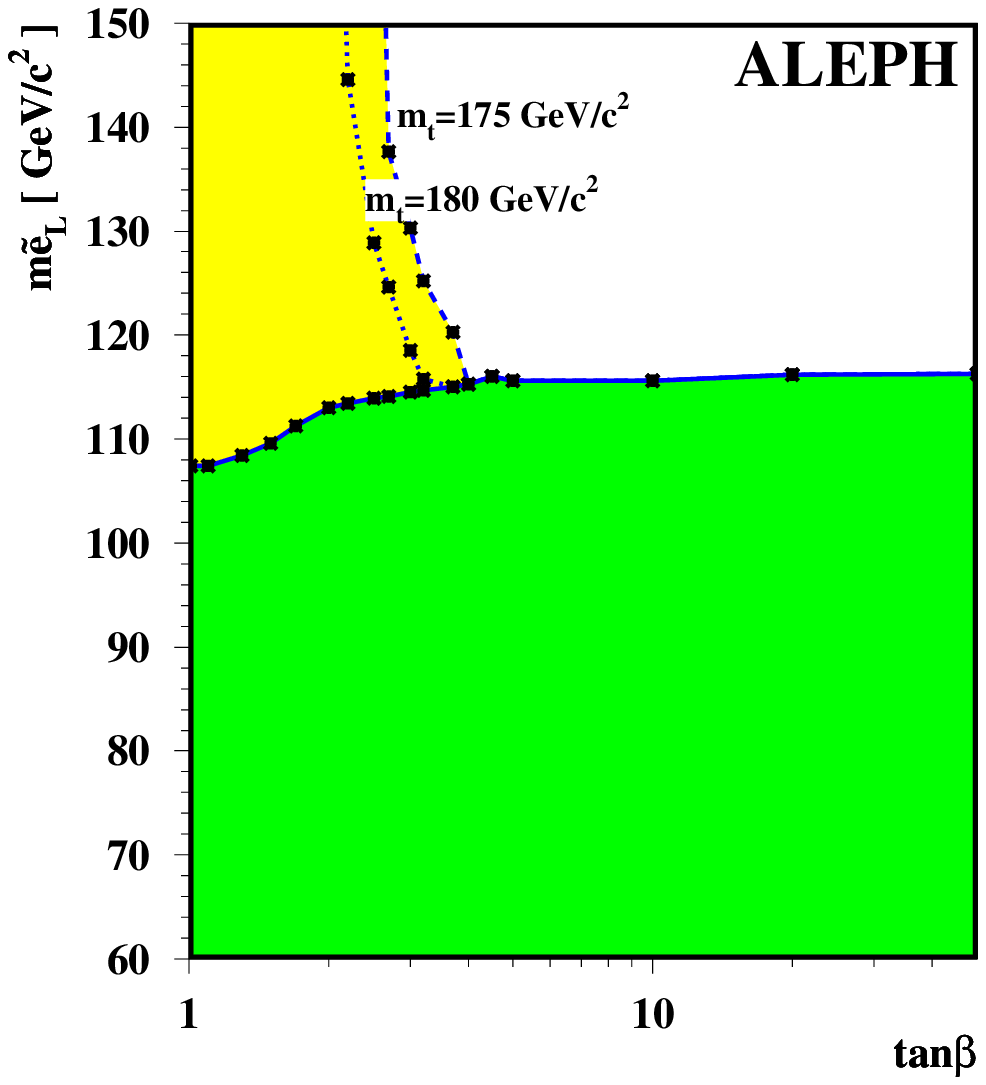,width=5cm,height=5cm}}
{\psfig{figure=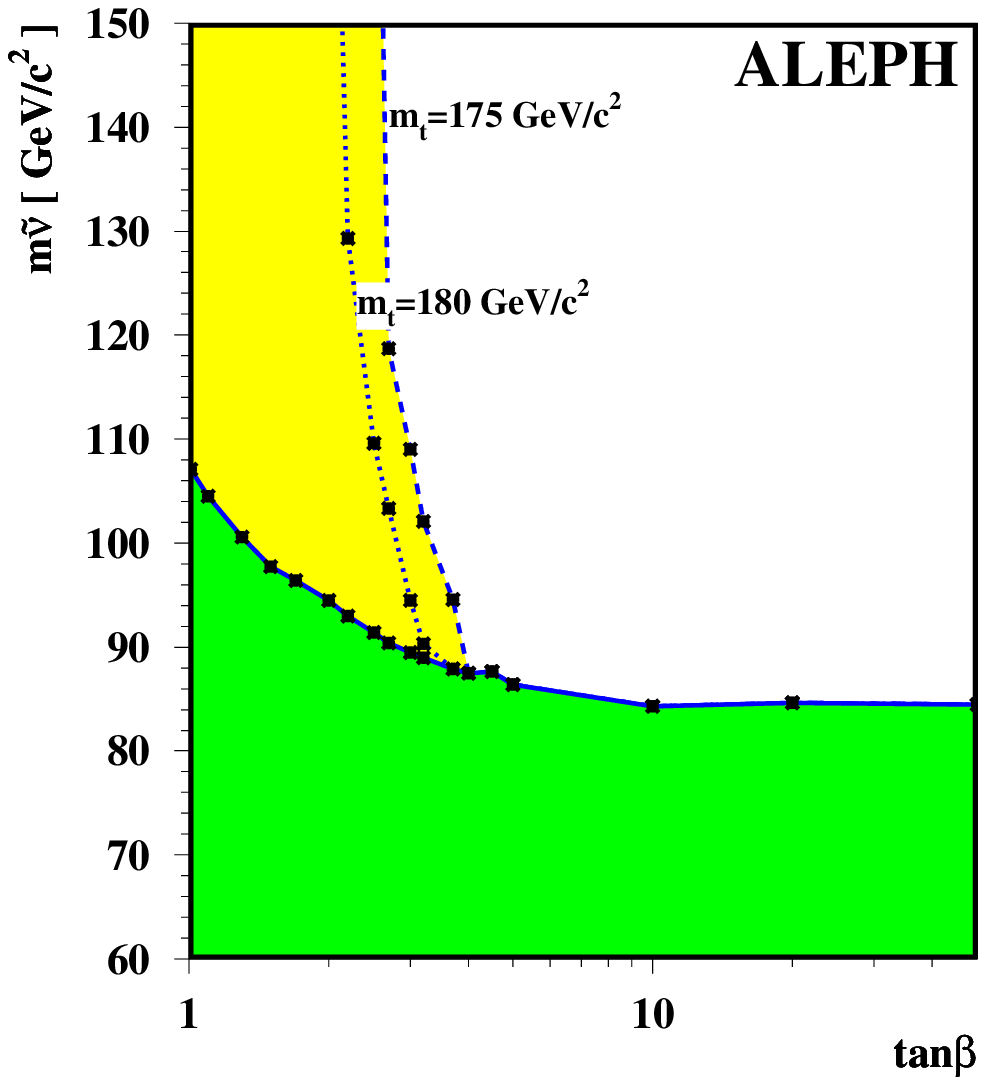,width=5cm,height=5cm}}

\end{center}
\caption{Mass exclusion plots versus $\tan \beta$ for   1) $\tilde e_R$ mass. 2) $\tilde e_L$ mass.  3) S $\tilde \nu $ mass. The influence of Higgs negative search is also added.
 \label{fig:selectron}}
\end{figure}

Adding  information from single electron analysis, the scalar
electron exclusion can be extended. Figure~\ref{fig:selectron} shows
preliminary results from ALEPH \cite{lepsusy} using both single and two acoplanar
electrons searches. Right selectron masses lower than $73 \, GeV$
are excluded at $ 95 \% $CL
 .  When unification relations are assumed, negative
searches of right selecton as well as those in the gaugino sector
can be translated into exclusion in the SUSY parameters  space
$m_0, m_{1/2}, tan \beta$. This can be then used to set limits on
left selectron and sneutrino masses. In
this way ALEPH excludes left selectron masses lower  than $107 \,
GeV $ and sneutrino masses lower than $ 83 \, GeV$ at $ 95 \% $CL .  In addition,
negative standard Higgs searches can be interpreted within SUSY
context and translated into exclusion on $tan \beta$ leading to
increase the previous  mass limits by few GeV  as shown on the same
figure~\ref{fig:selectron}.

\subsection{SQUARKS}
\begin{figure}

\begin{center}
{\psfig{figure=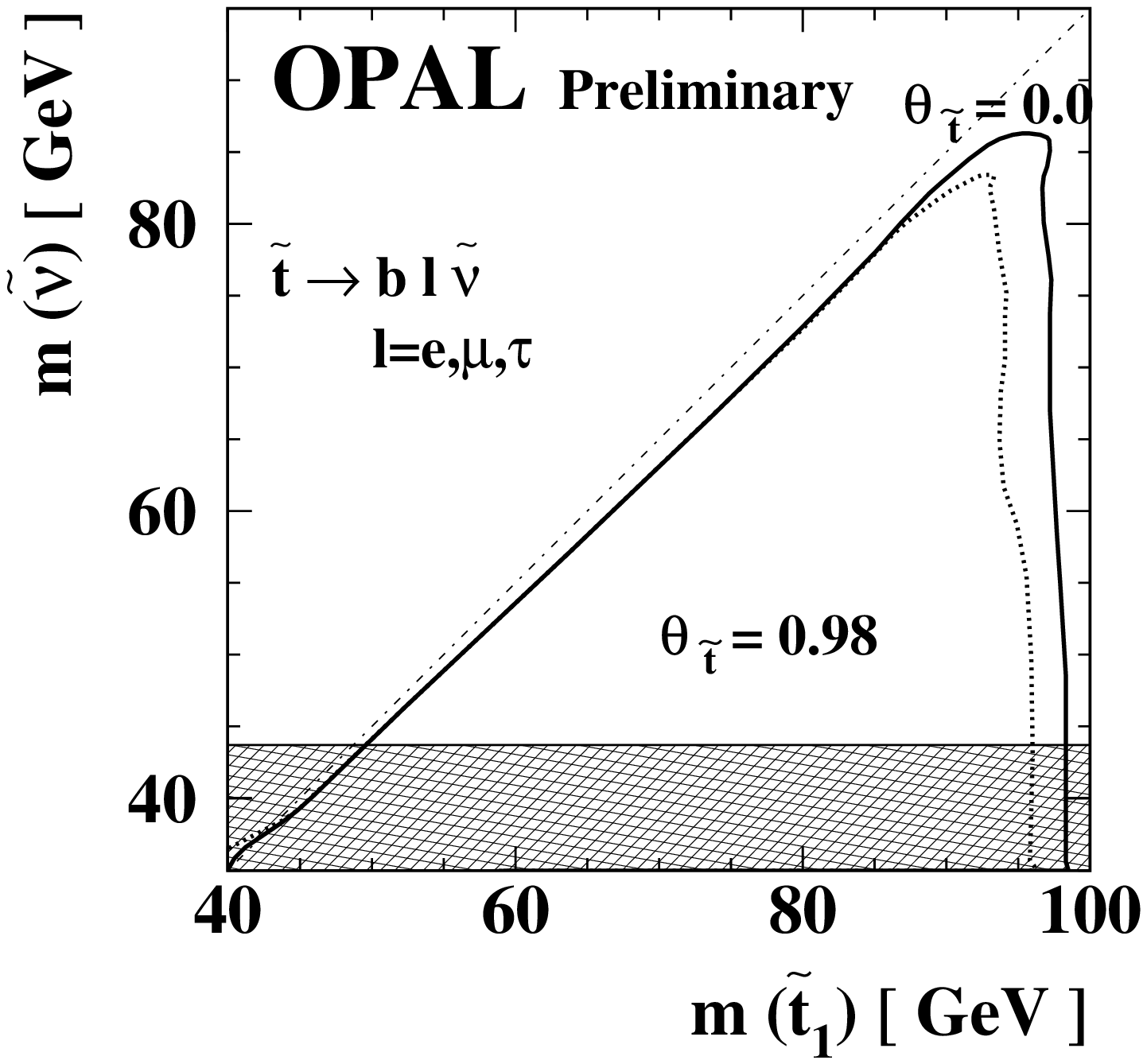,width=5cm,height=5cm}}
{\psfig{figure=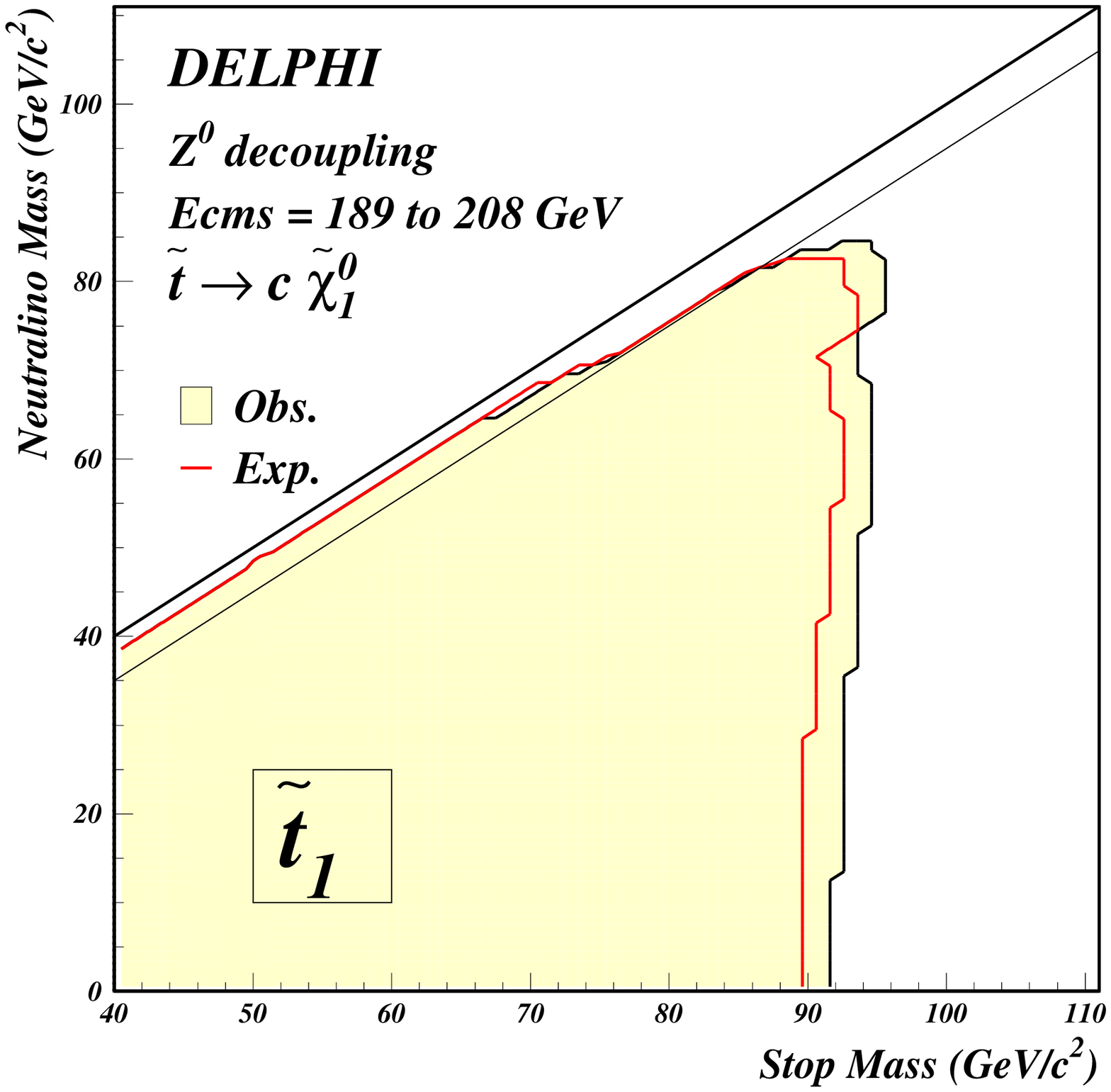,width=5cm,height=5cm}}
{\psfig{figure=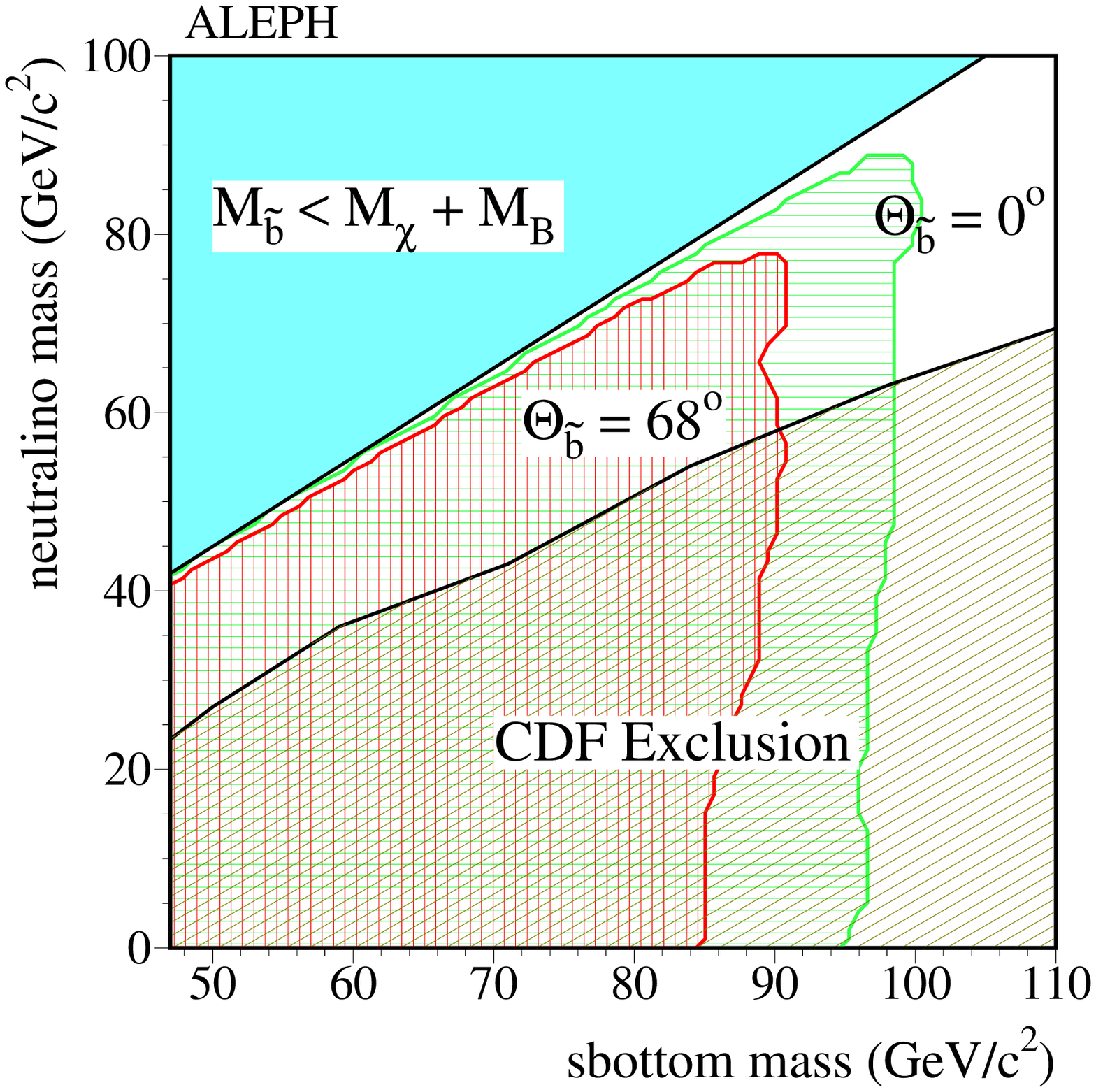 ,width=5cm,height=5cm}}
\end{center}

\caption{1)  $\tilde t_1$ mass exclusion area using $ \tilde t_1 \to b l \tilde \nu$ from OPAL. 2) $\tilde t_1$ mass exclusion area using  $ \tilde t_1 \to c \chi^0_1$ from  DELPHI.  \, 3) $\tilde b_1$  mass exclusion area from ALEPH.
\label{fig:stop}}

\end{figure}

Squarks are expected to be  heavier  than sleptons because of
their  additional interaction through QCD with
gluinos\footnote{Gluinos are expected to be heavier than the other
gauginos within the MSSM.}. Among the different scalar quarks, stop
$\tilde t_1$ and sbottom $\tilde b_1$ have been intensely studied
at LEP. This is related to the fact that under mixing hypothesis,
those two squarks can be light enough  to be produced at LEP.
\vglue .3 cm
\noindent
{\bf STOP:}
 $\tilde t_1$  pair production has been searched within many
scenarios corresponding to its relative mass with respect
to other SUSY particles it may decay in.  When $m(\tilde t_1)-m(\tilde \nu)) > m(b)\approx  5 \, GeV$, the $\tilde t_1$ decays principally through:
$ \tilde t_1 \to b l \tilde \nu$.  In this case the stop pair
production is searched by selecting events with two jets, two
leptons and missing energy since the sneutrino decays into a
neutralino and a neutrino both escaping detection.
Negative results within this scenario is shown by OPAL \cite{lepsusy} as an
exclusion contour on the $\tilde t_1$ mass in figure~\ref{fig:stop}.
When the previous channel is  kinematically forbidden, stop decays
through   the less favored two  FCNC processes: $\tilde t_1 \to u
\chi^0_1 ,\,\tilde t_1 \to c \chi^0_1.$ The second  is dominant when
the mass difference $\Delta m =  m(\tilde t_1)-m(\chi^0_1) > m(c) \approx 1.5 \,GeV$ and absent otherwise.  The nature
of the decay in the two previous cases leads probably $\tilde t_1$
to hadronize before decaying. This gives birth to different
experimental signatures. Systematic studies with different
hadronization scenarios were performed  by the different LEP
experiments. Absence of new signature is then transformed into an
exclusion area in the plane $(m(\tilde t_1),m(\chi^0_1))$
  as shown by
 DELPHI \cite{lepsusy} in figure~\ref{fig:stop} with the conservative choice of  a mixing angle of $\theta_{\tilde t} = 56^\circ$ corresponding to the decoupling scenario.  New stop decay channel namely $\tilde t_1 \to f \bar f'  \chi^0_1$ proceeding through 4-body decay was recently studied by ALEPH  in order to increase the sensitivity in the corridor region where $\tilde t_1$ is mass degenerate not only with $\chi^0_1$ but also with $ \chi_1^\pm$.
\vglue .3 cm
\noindent
{\bf SBOTTOM:} The main decay process of $\tilde b_1$ when the mass difference
$\Delta m =  m(\tilde b_1)-m(\chi^0_1) > m(b)$,  is
$\tilde b_1 \to b  \chi^0_1$ The sbottom pair production can be
therefore  investigated by looking for events with two acoplanar
b-tagged jets and  missing energy.  The  LEP experiments reported
no excess in this channel setting as previously  an exclusion area
in the plane $(m(\tilde b_1),m(\chi^0_1))$ corresponding to the
decoupling scenario with a mixing angle of $\theta_{\tilde b} =
68^\circ$ as shown by ALEPH in figure~\ref{fig:stop}.

\section{GAUGINO SEARCHES}
In SUSY the gaugino sector is made of charginos $ \chi^\pm_1,
\chi^\pm_2$ and neutralinos $ \chi^0_i (i=1,4)$. The first are
charged mass eigenstates obtained as  linear combinations of
supersymmetric partners of $W^\pm$  bosons  and the charged
higgsinos. The neutralinos are neutral mass eigenstates. They are
linear combinations of supersymmetric partner of the photon, the
$Z$ boson and the two neutral higgsinos. Higgsinos are
supersymmetric partners of the Higgs  two doublet field
components. Charginos and neutralinos are called gaugino-like(higgsino-like) when the gaugino(higgsino) components are larger
than the higgsino(gaugino) ones respectively.

\subsection{CHARGINOS}
Only the lightest chargino $ \chi^\pm_1$ is likely to be produced
at LEP. Search strategy  of this chargino is determined according
to the mass difference  $\Delta m = m(\chi^\pm_1)-m(\chi^0_1)$. At
high values of  $\Delta m  (> 4 \,GeV)$  the chargino decays
immediately after its production  whereas  at low  $\Delta m ( < 4 \,GeV)$ its decay can be delayed. This leads to different
topologies when looking for charginos:

\vglue .3 cm
\noindent
{\bf 1-High  $\Delta m$ chargino searches:}\\
 Each of the two charginos produced at the primary vertex decays
either hadronically $ \chi^\pm_1 \to \chi^0_1 q \bar q'$  or
leptonically $ \chi^\pm_1 \to \chi^0_1 l \nu$.  Charginos
production has therefore three kinds of topology: jets,
jets+leptons and only leptons. All of these  topologies have been
studied by the LEP experiments. The good agreement betwen data and
 the SM prediction in the four LEP experiments  allows to
constrain the SUSY parameters involved in the chargino production.
The chargino pair production proceeds through s and t-channel. The
t-channel contribution leads to a destructive interference and
hence to a decrease  of the cross-section. This decrease can be
very high for low $ \tilde \nu$ masses\footnote{The t-channel proceeds
through the exchange of a sneutrino.}  and negligible  for high ones.
Collecting the four LEP experiments chargino search results for
$(\Delta m > 4 \,GeV)$, the SUSY working group at LEP set a
preliminary lower limit on the chargino mass of $103.5 \,GeV$ in
case of sneutrino mass exceeding $300 \,GeV $ \cite{suwogr}. The
negative results in the neutralino sector can also be used to
increase the lower limit on the chargino mass since it constrains
the SUSY parameters and in some regions this leads to a chargino mass
excluded up to more than $6 \, GeV $ beyond the kinematic
limit as shown by ALEPH \cite{lepsusy}. 
\vglue .3 cm \noindent {\bf 2-Low
$\Delta m$ chargino searches:} \\ Three scenarios  are essentially
investigated by the LEP experiments. They depend on the value of
$\Delta m$ and on the chargino decay length:
\\
\noindent {\bf a- Quasi-stable charginos topology:} This scenario
occurs when the mass difference is lower than the pion mass. In
this case the two charginos may not decay inside the detector,
  giving rise to two stable heavy   muon-like particles. Absence of excess of this kind of
events has been translated into chargino mass limit at very low
$\Delta m$ as shown by OPAL\cite{lepsusy}. \\
\noindent {\bf b- Kink and secondary vertices
topology:}  Here the $\Delta m$ is large enough to allow the
chargino decay inside the detector but not at the primary vertex.
This gives birth in  large TPC detectors as those of DELPHI and
ALEPH to events with kink. Systematic study of this kind of events
has been done but no excess is observed.
\\
\noindent {\bf c-ISR topology:} In this case the chargino decays
promptly. However the decay products are too soft to be detected
and the event may not be triggered. Using an initial state
radiation photon (ISR) can overcome this difficulty when the
photon energy is large enough to trigger the event. A drawback of
this technique is the low number of events due to the
ISR requirement. Another difficulty is related to  the main
background contribution to this SUSY scenario which is the
2-photon physics not well simulated  in this domain of very low
energy. This increases the systematic uncertainty related  to this
scenario.  The nature of the chargino is important here.
Higgsino-like chargino with low
 $\Delta m$  is natural in the  constrained MSSM model whereas for gaugino-like
 one relation between $M_1$ and $M_2$ should be relaxed. No excess was
observed for this topology.

\begin{figure}

\begin{center}

{\psfig{figure= 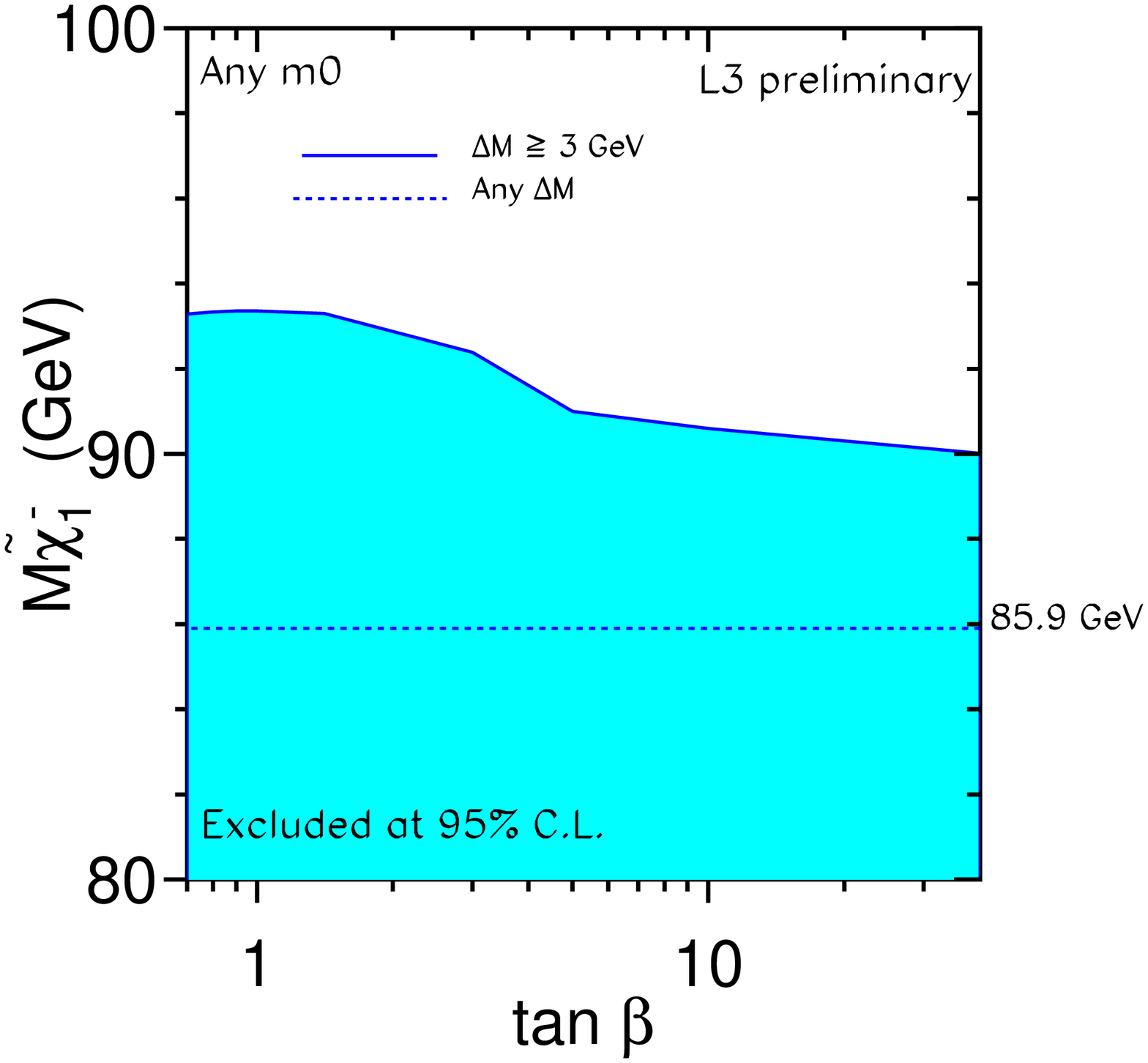,width=7cm,height=5cm}}
{\psfig{figure= 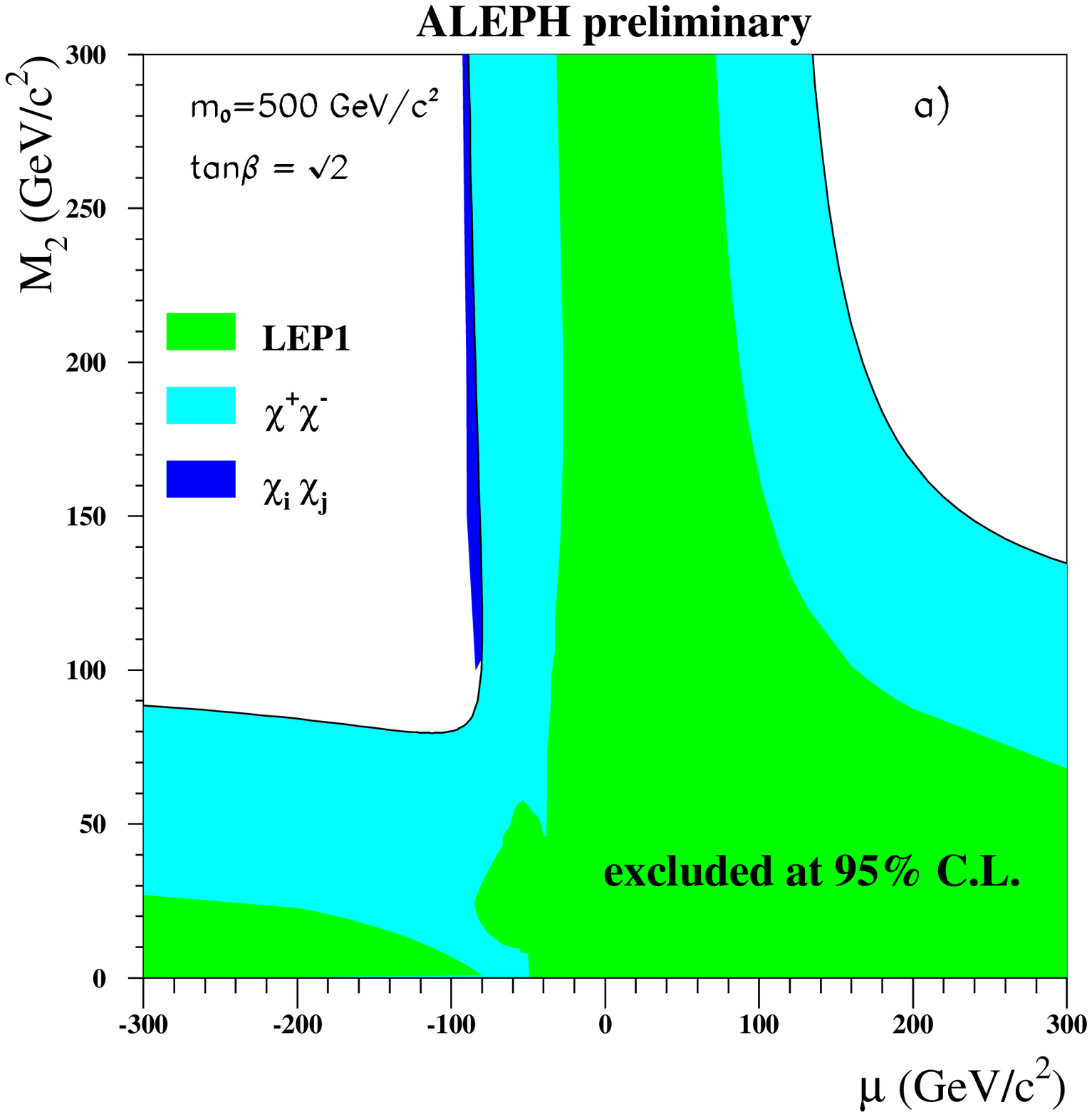,width=7cm,height=5cm}}

\end{center}

\caption{1)  Preliminary absolute mass limit on chargino mass from L3. 2) Exclusion region in the plane $(M_2,\mu)$ set by ALEPH using chargino and neutralino negative searches. \label{fig:charginol3}}

\end{figure}

Combining results for both low and high  $\Delta m$ leads to set
an absolute  lower limit on the chargino mass within the CMSSM as
stated by L3 \cite{lepsusy} experiment which excludes the chargino mass up to
$85.9 \, GeV$ at $95$ \% CL as shown in figure~\ref{fig:charginol3}.

\subsection{NEUTRALINOS}

The neutralino sector is very rich due to  the
presence of four  different neutralinos. Almost all the
combinations of neutralino pair production including the
different decaying scenarios have been considered by the  LEP
experiments. Still, the most interesting one is the $\chi^0_1
\chi^0_2$ production because it is the most probable in term  of
accessibility at LEP since the $\chi^0_1 \chi^0_1$ goes
undetected\footnote{ Using ISR $\gamma$ is helpless because of the
very low  cross-section of this process.}. The various scenarios
of $\chi^0_2$ decay are studied. They include hadronic, leptonic
and even radiative decay\footnote{
Radiative decay may occur when the two neutralinos
$\chi^0_2,\chi^0_1 $ are of opposite natures.} {$\chi^0_2 \to \chi^0_1 \gamma$}. The other pair productions  like
$\chi^0_i,\chi^0_j$~$(i,j>2)$  with multi-jet and mutli-lepton
as well as mixed leptons-jets final states have also been
considered. Events with tau cascades  resulting from  $\chi^0_2$
decay: $\chi^0_2 \to \tilde \tau \tau $ have been studied
carefully due to their importance in the LSP search as will be
explained later. In all these channels the data show no
significant deviation from the expected contribution  of the
Standard Model. The neutralino and chargino negative results are
summed up to constrain the SUSY parameter space reducing
considerably their  domain. Figure~\ref{fig:charginol3} shows how
negative results from ALEPH are translated into exclusion region
in the $(M_2,\mu)$ plane.
\section {LSP}
\begin{figure}

\begin{center}
{\psfig{figure=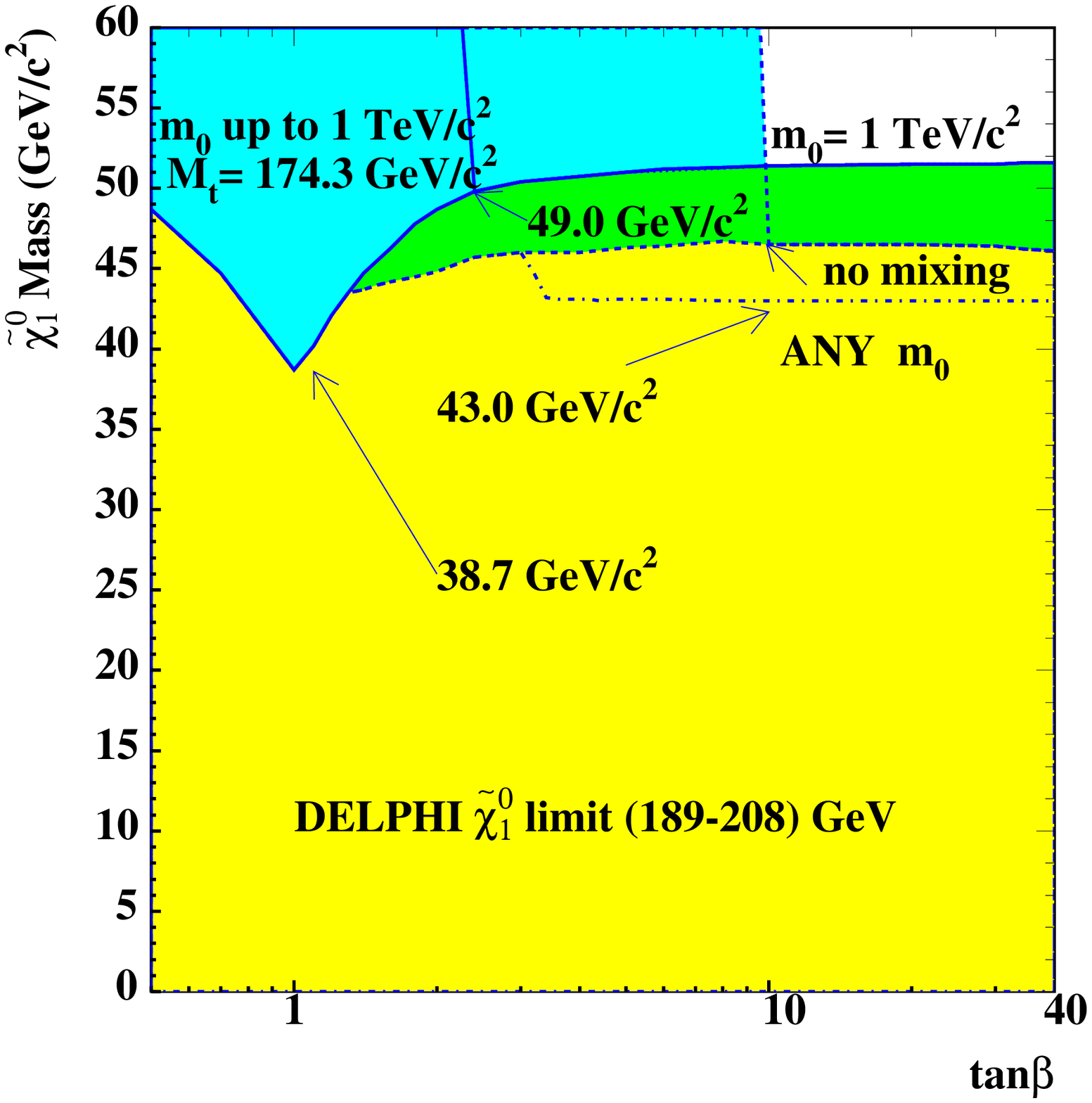,width=7cm,height=5cm}}
{\psfig{figure=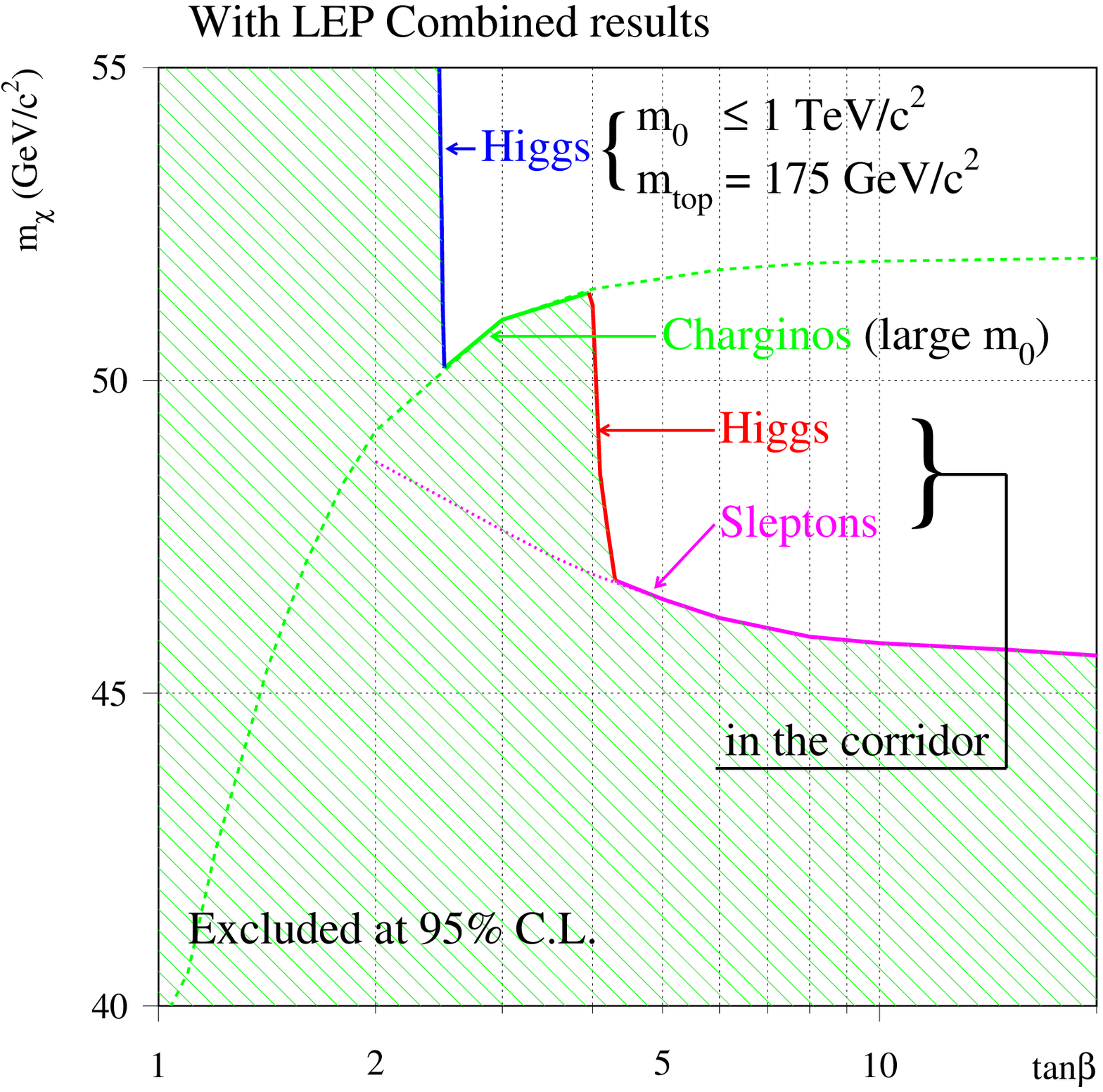,width=7cm,height=5cm}}

\end{center}

\caption{1) Preliminary LSP lower limit as function of $\tan \beta$ established by DELPHI. 2) LSP lower limit set by SUSY working group within the MSSM model. \label{fig:delphilsp}}

\end{figure}

So far the SUSY searches within the MSSM framework have been
sterile. The accumulated negative results either in the sfermion
sector or in the gaugino one allow to constrain  the SUSY parameters
space. Since the  LSP is the lightest neutralino for almost the
entire SUSY parameter space, the lower limit that will be set on
its mass will be the same for the LSP.   $\chi^0_1$  mass depends
on $M_2,  tan \beta, \mu$ and hence constraints on these parameters
will lead to a constraint on the LSP mass. This is achieved by
determining for different values of $tan \beta$, the exclusion
region in the plane $(M_2,\mu)$ and then setting a lower limit on
the LSP mass determined by these three parameters. Unfortunately,
determining the exclusion zone in the plane $(M_2,\mu)$ for a fixed
value of $\tan \beta $ is not straightforward. It should take into
account the different values of the other SUSY parameters ($m_0$
, $A$)
  since they affect branching ratios and masses of
sfermions and therefore may change dramatically the SUSY events
topology. For example, low value of $m_0$ will lead to small mass
sneutrino reducing the exclusion region from chargino negative
searches. This scenario can become even worse when the sneutrino
is slightly heavier than a chargino degenerate in mass with
the lightest neutralino (the so-called corridor problem). $A$ parameter
effect can be also  very important since it affects $\tilde \tau$,
$\tilde t$, and $\tilde b$ masses. If $A_\tau$ is such that
$\tilde \tau$ is almost degenerate in mass with $\chi^0_1$, search
for  two acoplanar taus can become inefficient and needs to be
completed by taus cascade configurations resulting from
$\chi^0_2$ decay as mentioned in the previous section. These
complications have pushed to use negative SUSY search results of
all the  possible decay scenarios in order to cover as
much as possible  those  inaccessible regions. Considering all
these configurations, the  lower
LSP mass limit may be expressed as a function of  $\tan \beta$  with an absolute
 minimum obtained for $\tan \beta = 1$  as can be shown from the
preliminary result from  DELPHI in figure~\ref{fig:delphilsp} where
different values of $m_0$ and $A$ are considered. Negative
standard Higgs search at LEP was interpreted in the frame of  the MSSM
Higgs sector leading to exclude  low  $\tan \beta$ values. This
shifts upward the LSP lowest value by few GeV  depending on the top
mass. Indeed the Higgs exclusion analysis depends strongly on the
top mass value through radiative corrections. Increasing $m(top)$
from 175 to 180  $ GeV$ results in reducing the LSP mass limit
by about $1 \,GeV$. SUSY working group at LEP has started to
merge  results from the four experiments to set a LEP limit on
the LSP. A preliminary result is presented in figure~\ref{fig:delphilsp} where the
Higgs negative results are used in a conservative
way\footnote{using ALEPH results which represents the lowest limit
on Higgs}. The limit on the LSP is about $ 45 \, GeV$
 at $ 95$ \% CL
established at high values of $\tan \beta$. Using Msugra, the same
group has set a lower limit of $52.2 \, GeV$ in the absence of the
trilinear coupling ($A=0$).

 \section{CONCLUSION}

The LEP experiments have been looking for SUSY particles up to the
kinematic limits for most of them. Many  MSSM  scenarios have
been investigated but no new physics manifestation is observed.
Preliminary limits on parameters, cross-sections and particles
masses have been set and much more precise results will come soon.
An important result is the exclusion of the LSP mass up to more than
$40 \, GeV$ without including the negative Higgs results. This is
a huge improvement of our knowledge with respect to the situation
before LEPII.

\section*{Acknowledgments}
I would like to thank  J.J.Blaising, M.Chemarin, G.Coignet, J.Fay, J.P Martin
 and S.Rosier-Lees for useful discussions in preparing this conference.


\section*{References}
\begin{thebibliography}{99}
\bibitem{susy} For a SUSY review: H.P.Nilles,''Supersymmetry, Supergravity And Particles Physics'', Phys. Rep. 100(1984)1; H.E.Haber and G.L.Kane, ``The Search For Supersymmetry: Probing Physics Beyond The Standard Model'', Phys.Rep 117(1985)75.
\bibitem{RGE} L.E Ib\'a\~neza, C.Lopez and C.Mu\~noz, Nucl.Phys. B256(1985)218.
\bibitem{suwogr} http//lepsusy.web.cern.web.ch/lepsusy.
\bibitem{lepsusy} All the preliminary results quoted in this paper can be obtained from the four LEP experiments web pages accessible from: http://greybook.cern.ch.
\end{thebibliography}
\end{document}